# Simultaneous test for Means: An Unblind Way to the F-test in One-way Analysis of Variance

Elsayed A. H. Elamir[1]

Management & Marketing Department, College of Business, University of Bahrain, P.O. Box 32038, Kingdom of Bahrain

## Abstract

After rejecting the null hypothesis in the analysis of variance, the next step is to make the pairwise comparisons to find out differences in means. The purpose of this paper is threefold. The foremost aim is to suggest expression for calculating decision limit that enables us to collect the test and pairwise comparisons in one step. This expression is proposed as the ratio of between square for each treatment and within sum of squares for all treatments. The second aim is to obtain the sampling distribution of the proposed ratio under the null hypothesis. This sampling distribution is derived exactly as the beta distribution of the second type. The third aim is to use beta distribution of second type and adjusted p-values to create adjusted points and decision limit. Therefore, reject the null hypothesis of equal means if any adjusted point falls outside the decision limit. Simulation study is conducted to compute type I error. The results show that the proposed method controls the type I error near the nominal values using Benjamini-Hochberg'adjusted p-values. Two applications are given to show the benefits of the proposed method.

***Keywords:*** ANOVA, adjusted p-value, beta distribution; Bonferonni's approximation, F-test, linear models.

## 1 Introduction

Fisher (1918) discussed the term variance and introduced analysis of variance that becomes well known after being included in Fisher's book (1925). Analysis of variance (ANOVA) is a collection of statistical models that used to analyze the differences among group means and overall mean where the sample variance is partitioned into components attributable to different sources of variation. ANOVA models are multilateral statistical tools for studying the relation between a response variable and one or more explanatory or predictor variables. These models do not need any assumptions about the nature of the statistical relation between the response and explanatory variables, nor do they need that explanatory variables to be quantitative; see, for example, Cochran and Cox (1957), Kutner et al. (2004), Montgomery and Runger (2011), Elamir (2012) and Montgomery (2013).

---

[1]Email: shabib@uob.edu.bh, shabib40@gmail.com

A simultaneous test for means called gANOVA is proposed based on the ratio of between square for each treatment and within sum of squares for all treatments. This ratio is created from F-test in one-way analysis of variance. This ratio is considered as two independent gamma random variables. The exact sampling distribution of this ratio under the null hypothesis is derived exactly as the beta distribution of the second type. An upper decision limit is obtained using adjusted p-value and beta distribution of second type to graph this ratio and reject the null hypothesis if any point falls outside the decision limit. The adjusted p-values are obtained using several methods. One of these methods is the Benjamini and Hochberg (1995) method that depends on the concept of false discover rate and gives the best result among other methods. However, gANOVA is not intended to replace ANOVA but to gives more explanation and analysis for differences among group means. Moreover, gANOVA could be considered as an unblind way for F-test to determine which specific group mean(s) is different from overall mean simultaneously or graphically. Simulation study is conducted to compute type I error for gANOVA. The results show that gANOVA based on adjusted p-values using Benjamini and Hochberg's method controls the type I error very well.

Two applications are given to show the benefits of the proposed method. In the first application the method is explained and applied to photosynthetic rates of the oak seedlings data. In the second application the method is applied to simulated data to show another benefit of the proposed method.

The fixed effect model is reviewed in Section 2. gANOVA and its sampling distribution are derived in Section 3. Two applications are studied in Section 4. Section 5 is devoted to conclusion. R-program is given in Appendix.

## 2 Single-Factor ANOVA model

Assume that there are $G$ different groups with individuals in each group $Y_{gi}$, $i = 1,2,...,n_g$, $n_T = n_1 + \cdots + n_G$ and $g = 1,...,G$. Let $Y_{gi} - \bar{Y}$ is the total deviation ($\bar{Y} = \sum_g^G \sum_i^{n_g} Y_{gi}/n_T$), $\bar{Y}_g - \bar{Y}$ is the deviation of grouped mean ($\bar{Y}_g = \sum_{i=1}^{n_g} Y_{gi}/n_g$) around overall mean, and $Y_{gi} - \bar{Y}_g$ is the deviation of individuals around the grouped mean. It is useful to describe the observations from an experiment with a model. The means model can be written as

$$Y_{gi} = \mu_g + \varepsilon_{gi}$$

$Y_{gi}$ is the value of the response variable in $i$th trial for the $g$th treatment, $\mu_g$ are parameters, $\varepsilon_{gi}$ are independent identically distributed normal with $N(0, \sigma^2)$; see, Kutner et al. (2004) and Montgomery (2013). The appropriate hypotheses are

$$H_0: \mu_1 = \mu_2 = \cdots = \mu_G$$
$$H_1: \mu_g \neq \mu_j \text{ for at least one pair}$$

The name analysis of variance is obtained from a partition of total variability into its component parts. The total corrected sum of squares

$$SST = \sum_{g=1}^{G} \sum_{i=1}^{n_g} (Y_{gi} - \bar{Y})^2$$

is used as a measure of overall variability in the data. Note that the total corrected sum of squares $SST$ may be written as

$$SST = \sum_{g=1}^{G} n_g (\bar{Y}_g - \bar{Y})^2 + \sum_{g=1}^{G} \sum_{i=1}^{n_g} (Y_{gi} - \bar{Y}_g)^2 = SSTR + SSE$$

Where $SSTR$ is called the treatment sum of squares (i.e., between treatments) and $SSE$ is called the error sum of squares (i.e., within treatments). Specifically, the mean squares for the treatment can be written as

$$MSTR = SSTR/(G-1) = \sum_{g=1}^{G} n_g (\bar{Y}_g - \bar{Y})^2 / (G-1)$$

is an estimate of $\sigma^2$ if the treatment means are equal. Also, the mean squares error is

$$MSE = \sum_{g=1}^{G} \sum_{i=1}^{n_g} (Y_{gi} - \bar{Y}_g)^2 / (n_T - G)$$

is a pooled estimate of the common variance $\sigma^2$ within each of the $G$ treatments. The expected value of $MSE$ is

$$E(MSE) = \sigma^2$$

and the expected value for $MSTR$ is

$$E(MSTR) = \sigma^2 + \frac{\sum_{g=1}^{G} n_g \tau_g^2}{G-1}$$

Therefore, if treatment means do differ, the expected value of the treatment mean square is greater than $\sigma^2$; see, for example, Montgomery (2013) and Kutner et al. (2004).

Therefore, if the null hypothesis of no difference in treatment means is true, the ratio

$$F_0 = \frac{MSTR}{MSE} = \frac{\sum_{g=1}^{G} n_g (\bar{Y}_g - \bar{Y})^2 / (G-1)}{\sum_{g=1}^{G} \sum_{i=1}^{n_g} (Y_{gi} - \bar{Y}_g)^2 / (n_T - G)}$$

is distributed as $F$ with $G-1$ and $n_T - G$ degrees of freedom. In practice it can conclude that there are differences in the treatment means if

$$F_0 > F_{\alpha; G-1, n_T-G}$$

where $F_0$ is the computed value and the distribution of $F_0$ is just the ratio of two independent gamma random variables

$$F_0 \sim \frac{\chi^2(G-1)/(G-1)}{\chi^2(n_T-G)/(n_T-G)} \sim \frac{\text{gamma}\left(\frac{G-1}{2}, \frac{G-1}{2}\right)}{\text{gamma}\left(\frac{n_T-G}{2}, \frac{n_T-G}{2}\right)} \sim F(G-1, n_T-G)$$

This is a special case of beta distribution of the second type; see, Coelho and Mexia (2007) and Garcia and Jaimez (2010).

## 3  Simultaneous test for means (gANOVA)

The computed $F_0$ can be written as

$$F_0 = \sum_{g=1}^{G} \left[ \frac{n_g (\bar{Y}_g - \bar{Y})^2 / (G-1)}{\sum_{g=1}^{G} \sum_{i=1}^{n_g} (Y_{gi} - \bar{Y}_g)^2 / (n_T - G)} \right] = \sum_{g=1}^{G} K_g$$

Hence,

$$F_0 = K_1 + K_2 + \cdots + K_G$$

where

$$K_g = \frac{n_g(\bar{Y}_g - \bar{Y})^2/(G-1)}{\sum_{g=1}^{G}\sum_{i=1}^{n_g}(Y_{gi} - \bar{Y}_g)^2/(n_T - G)}, \quad g = 1, 2, \ldots, G$$

is the ratio of between square for each treatment and within sum squares for all treatments.

Under the assumptions of

(a) fixed effect model and

(b) $(\bar{Y}_g - \bar{Y})^2$ and $\sum_{g=1}^{G}\sum_{i=1}^{n_g}(Y_{gi} - \bar{Y}_g)^2$ are gamma independently random variables.

Therefore, if the null hypothesis of no differences in treatment means is true, hence,

$$\sum_{g=1}^{G}\sum_{i=1}^{n_g}(Y_{gi} - \bar{Y}_g)^2/(n_T - G) \sim \sigma^2 \chi^2(n_T - G)/(n_T - G)$$

and

$$\frac{n_g(\bar{Y}_g - \bar{Y})^2}{G-1} = \frac{n_g(\bar{Y}_g - \mu)^2 - n_g(\bar{Y} - \mu)^2}{G-1} \sim \sigma^2 \frac{(n_T - n_g)}{(G-1)n_T}\chi^2(1)$$

The sampling distribution of $K_g$ can be expressed as

$$K_g \sim \frac{(n_T - n_g)\chi^2(1)/(G-1)n_T}{\chi^2(n_T - G)/(n_T - G)}$$

Hence,

$$K_g = \frac{n_g(\bar{Y}_g - \bar{Y})^2/(G-1)}{\sum_{g=1}^{G}\sum_{i=1}^{n_g}(Y_{gi} - \bar{Y}_g)^2/(n_T - G)} \sim \frac{(n_T - n_g)\chi^2(1)/((G-1)n_T)}{\chi^2(n_T - G)/(n_T - G)} \sim \frac{\text{gamma}\left(\frac{1}{2}, \frac{n_T(G-1)}{2(n_T - n_g)}\right)}{\text{gamma}\left(\frac{(n_T - G)}{2}, \frac{(n_T - G)}{2}\right)}$$

The exact sampling distribution for $K_g$ is given in the following theorem.

**Theorem**

Under the assumptions (a) and (b), the exact sampling distribution of $K_g$ is

$$f_{K_g}(k) = \frac{\left[(n_T(G-1))/\left((n_T - n_g)(n_T - G)\right)\right]^{-1/2}}{B\left(\frac{1}{2}, \frac{n_T - G}{2}\right)}\left(1 + \frac{n_T(G-1)}{(n_T - n_g)(n_T - G)}k\right)^{-(n_T - G + 1)/2} k^{-1/2},$$

$k > 0, g = 1, \ldots, G$

This distribution is defined in terms of $G$, $n_g$ and $n_T$ and is a special case from beta distribution from the second type.

*Proof*

Coelho and Mexia (2007) have given the distribution of the ratio of two independent random variables

$$Z = Y_1/Y_2$$

each has gamma distribution as

$$f_Y(y) = \frac{\lambda^r}{\Gamma(r)} y^{r-1} e^{-\lambda y}, \quad r, \lambda, y > 0$$

$\lambda$ is scale parameter and $r$ is the shape.

They have given the distribution of $Z$ as

$$f_Z(z) = \frac{\left(\frac{\lambda_1}{\lambda_2}\right)^{r_1}}{B(r_1, r_2)} \left(1 + \frac{\lambda_1}{\lambda_2} z\right)^{-r_1 - r_2} z^{r_1 - 1}, \quad z > 0$$

$B(.,.)$ is a beta function and this distribution is most commonly known as beta distribution of the second type (GB2). By putting

$$\lambda_1 = (n_T(G-1))/(2(n_T - n_g)), \quad \lambda_2 = (n_T - G)/2, \quad r_1 = 1/2, \quad \text{and} \quad r_2 = (n_T - G)/2$$

The sampling distribution of $K$ is obtained.

**Corollary**

Under the assumptions (a) and (b) and equal sample sizes in each group $n_1 = n_2 = \cdots = n_G = n$, the exact sampling distribution of $K_g$ is simplified to

$$f_{K_g}(k) = \frac{[1/(n-1)]^{-1/2}}{B\left(\frac{1}{2}, \frac{G(n-1)}{2}\right)} \left(1 + \frac{1}{(n-1)} k\right)^{-(G(n-1)+1)/2} k^{-1/2},$$

In the case of equal sample sizes, the non-central moments for $K$ distribution can be obtained from Coelho and Mexia (2007) as

$$E(K^j) = (n-1)^j \frac{\Gamma(0.5 + j)\Gamma\left(\frac{G(n-1)}{2} - j\right)}{\Gamma(0.5)\Gamma\left(\frac{G(n-1)}{2}\right)}$$

The first two moments can be obtained as

$$E(K) = (n-1) \frac{\Gamma(1.5)\Gamma\left(\frac{G(n-1)}{2} - 1\right)}{\Gamma(0.5)\Gamma\left(\frac{G(n-1)}{2}\right)}$$

and

$$V(K) = (n-1)^2 \left[ \frac{\Gamma(2.5)\Gamma\left(\frac{G(n-1)}{2} - 2\right)}{\Gamma(0.5)\Gamma\left(\frac{G(n-1)}{2}\right)} - \left[\frac{\Gamma(1.5)\Gamma\left(\frac{G(n-1)}{2} - 1\right)}{\Gamma(0.5)\Gamma\left(\frac{G(n-1)}{2}\right)}\right]^2 \right]$$

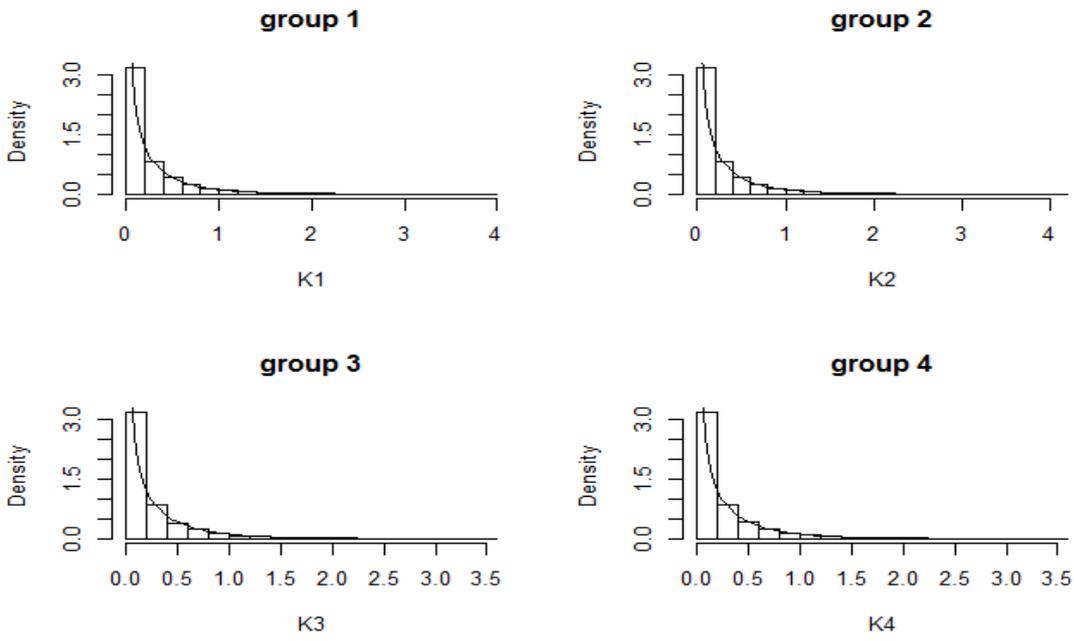

Figure 1. Histogram of $K_g$ ($G = 4$ and $n_g = 20$) using simulated data from normal distribution with beta distribution of second type superimposed.

Figure 1 shows the histogram and density for $K_1, \ldots, K_4$ using simulated data from normal distribution and $n_g = n = 20$ for each group. The distribution gives a very good fit for the simulated data.

## 3.1 Decision limit for gANOVA

Multiple testing refers to any instance that involves the simultaneous testing of more than one hypothesis. Failure to control Type I error when examining multiple outcomes may yield false inference. Several methods are based on the Bonferroni and Sidak inequalities (Sidak, 1967;

Simes, 1986) that maximize power while ensuring an acceptable Type I error rate. These methods adjust $\alpha$ values or $p$-values using simple functions of the number of tested hypotheses; see, Benjamini(2010), Bretz et al. (2011) and Westfall (2005). Holm (1979), Hochberg (1988), and Hommel (1988) developed Bonferroni derivatives incorporating stepwise components.

To find the decision limit for $K_g, g = 1,2,..,G$, there are multiple tests ($G$ tests) and it is needed to distinguish between two meanings of $\alpha$ when performing multiple tests:

1. The probability of making a Type I error when dealing only with a specific test. This probability is denoted $\alpha[PT]$ ("alpha per test"). It is also called the test-wise alpha.
2. The probability of making at least one Type I error for the whole family of tests. This probability is denoted $\alpha[PF]$ ("alpha per family of tests"). It is also called the family-wise or the experiment-wise alpha.

Dunn-Sidak-Bonferroni methods and their relatives are the standard approach for controlling the experiment-wise alpha by specifying what $\alpha$ values should be used for each individual test (i.e., the test is declared to be significant if $p \leq \alpha$). Hence, the probability of at least one Type I error for the whole family is

$$\alpha(PF) = 1 - (1 - \alpha(PT))^G$$

If one wishes the test-wise alpha for the independence tests, it can be obtained as

$$\alpha(PT) = 1 - (1 - \alpha(PF))^{1/G}$$

This is often called the Dunn-Sidak method, for more details; see, for example, Dunn (1964) and Abdi (2007). Noting that $(1 - \alpha)^{1/G} \cong 1 - (1/G)\alpha$, the Bonferroni approximation gives

$$\alpha(PT) \approx \frac{\alpha(PF)}{G}$$

For example, to perform four tests, $G = 4$, and the risk of making at least one Type I error to an overall value of $\alpha(PF) = 0.05$, a test reaches significance if its associated probability is smaller than or equal to

$$\alpha(PT) = 1 - (1 - 0.05)^{1/4} = 0.01274$$

using the Bonferonni approximation

$$\alpha(PT) \approx \frac{\alpha(PF)}{G} = \frac{0.05}{4} = 0.0125$$

Under a strict Bonferroni correction, only hypotheses with associated values less than or equal to $\alpha(PT)$ are rejected, all others are accepted. When the null hypothesis is rejected, the multiple comparison correction should take this into account. There are many methods such as Holm's method (1979), Simes-Hochberg method (Simes 1986, Hochberg 1988) and Hommel's method (1988 and 1989).

Another good method due to Benjamini and Hochberg (1995) that depends on the concept of false discover rate (FDR) that is designed to control the expected proportion of rejected null hypotheses that were incorrect rejections (false discoveries). Note that he Benjamini–Hochberg procedure (BH) controls the false discovery rate at level $\alpha$.

In any case the R-software has several methods under the function

p.adjust(p, method = "", n = length(p))

These methods are c("holm", "hochberg", "hommel", "bonferroni", "BH", "BY"); BH: Benjamini–Hochberg and BY: Benjamini and Yekutieli.

**Proposed gANOVA**

By using these methods, the gANOVA can be proposed as

$$H_0: \mu_1 = \mu_2 = \cdots = \mu_G$$

is rejected if

$$\text{any}(p_{adjust}) < \alpha(PF)$$

Graphically this can be shown using two methods that give the same conclusion.

Firstly, by putting

$g$ on $x$ axis versus $1 - p_{adjusted}$ on $y$ axis with $DL = 1 - \alpha(PF)$

and $H_0$ is rejected if

$$\text{Any}(1 - p_{adjusted}) > DL$$

Secondly by using the quantile function for the second type beta distribution where the decision limit can be proposed as the upper limit for the quantile of second type beta distribution at $(1 - \alpha(PF))$. The decision limit can be obtained using the quantile function of second beta distribution and R-software function (qGB2 from package GB2) as

$$DL = qGB2\left(1 - \alpha(PF), 1, \frac{(n_T - G)(n_T - n_g)}{n_T(G - 1)}, \frac{1}{2}, \frac{(n_T - G)}{2}\right)$$

and $K_{adjusted}$ can be also obtained from the second type beta distribution as

$$K_{adjusted} = qGB2\left(1 - p_{adjusted}, 1, \frac{(n_T - G)(n_T - n_g)}{n_T(G - 1)}, \frac{1}{2}, \frac{(n_T - G)}{2}\right)$$

and $H_0$ is rejected if

$$\text{Any}(K_{adjusted}) > DL$$

### 3.2 Simulation study

Simulation study is conducted using data from normal distribution as following.

1. Simulate data from normal distribution with means equal 0 for all groups and unit variance,
2. The number of groups $(G)$ is $3, 5, 10$ and $20$,
3. Sample size in each group $n_g = 10, 20, 50$ and $100$,
4. The nominal level of significance is $0.05$ and $0.01$.
5. Four methods are used. These methods are Bonferroni, Hommel, Benjamini-Hochberg and ANOVA.
6. The estimated level of significance is computed as the percentage of the number of rejected $H_0$ when $H_0$ is true.
7. The number of replications is 10000.
8. The average of estimated level of significance is computed and reported in Table 1.

The comparison among Bonferroni, Hommel, BH and ANOVA methods in terms of type one error (family-wise alpha) is given in Table 1. As it can be seen when the number of groups is small the BH method is the nearest method to nominal values (0.05 and 0.01) and ANOVA. When the number of groups becomes larger, all methods are very good in comparison with ANOVA and nominal values (0.05 and 0.01). From these results, the BH method is recommended to adjust p-values that used in building gANOVA.

Table 1. Empirical Type I error (family-wise) for $K_g$ using Bonferroni (Bonfe), Hommel (Homm), Benjamini–Hochberg (BH) and ANOVA methods based on simulated data from normal distribution and the number of replications is 10000.

|  | $\alpha = 0.05$ | | | | $\alpha = 0.01$ | | | |
| --- | --- | --- | --- | --- | --- | --- | --- | --- |
| | | | | $G = 3$ | | | | |
| $n_g$ | Bonfe | Homm | BH | ANOVA | Bonfe | Homm | BH | ANOVA |
| (all 10) | 0.0435 | 0.0448 | 0.0466 | 0.0515 | 0.0073 | 0.0075 | 0.0082 | 0.0086 |
| (all 20) | 0.0415 | 0.0429 | 0.0449 | 0.0452 | 0.0075 | 0.0078 | 0.0082 | 0.0082 |
| (all 50) | 0.0440 | 0.0467 | 0.0471 | 0.0500 | 0.0096 | 0.0099 | 0.0099 | 0.0101 |
| (all 100) | 0.0447 | 0.0462 | 0.0479 | 0.0504 | 0.0097 | 0.0099 | 0.0101 | 0.0103 |
| | | | | $G = 5$ | | | | |
| (all 10) | 0.0465 | 0.0480 | 0.0490 | 0.0490 | 0.0091 | 0.0091 | 0.0091 | 0.0091 |
| (all 20) | 0.0465 | 0.0469 | 0.0483 | 0.0489 | 0.0099 | 0.0099 | 0.0099 | 0.0105 |
| (all 50) | 0.0494 | 0.0502 | 0.0512 | 0.0518 | 0.0094 | 0.0095 | 0.0097 | 0.0090 |
| (all 100) | 0.0480 | 0.0482 | 0.0496 | 0.0516 | 0.0096 | 0.0097 | 0.0099 | 0.0107 |
| | | | | $G = 10$ | | | | |
| (all 10) | 0.0463 | 0.0465 | 0.0482 | 0.0491 | 0.0099 | 0.0099 | 0.0100 | 0.0092 |
| (all 20) | 0.0442 | 0.0445 | 0.0468 | 0.0481 | 0.0103 | 0.0103 | 0.0104 | 0.0096 |
| (all 50) | 0.0461 | 0.0465 | 0.0488 | 0.0494 | 0.0114 | 0.0114 | 0.0115 | 0.0115 |
| (all 100) | 0.0477 | 0.0480 | 0.0487 | 0.0488 | 0.0103 | 0.0103 | 0.0103 | 0.0096 |
| | | | | $G = 20$ | | | | |
| (all 10) | 0.0467 | 0.0468 | 0.0485 | 0.0490 | 0.0117 | 0.0117 | 0.0117 | 0.0116 |
| (all 20) | 0.0489 | 0.0491 | 0.0500 | 0.0524 | 0.0112 | 0.0112 | 0.0112 | 0.0090 |
| (all 50) | 0.0496 | 0.0497 | 0.0502 | 0.0550 | 0.0103 | 0.0103 | 0.0103 | 0.0092 |
| (all 100) | 0.0501 | 0.0502 | 0.0511 | 0.0515 | 0.0101 | 0.0101 | 0.0102 | 0.0097 |

## 4 Applications

The proposed method is applied to photosynthetic rates of the oak seedlings data and simulated data from normal distribution.

## 4.1 Photosynthetic rate of the oak seedlings

"In 2015 researchers planted several hundred oak seedlings in four horizontal transects at different elevations on a sandy ridge: one at the bottom, one at the top, and two more at equally spaced intervals in between. They anticipated that transect location might affect the photosynthetic rates of the oak seedlings because water availability in the soil declined with elevation…", see for more details and data at https://www.stthomas.edu/biology.

The null hypothesis for the test is that there are no differences in mean photosynthetic rate among the four groups of seedlings planted along each of the four transects.

The Shapiro normality test gives p-value 0.0000001 that indicates that the normality assumption is not suitable for photosynthetic data. Also, the Bartlett test of homogeneity of variances gives p-value 0.006 that does not support homogeneity of variances.

The photosynthetic rate has been square root transformed to improve the fit of the data to a normal distribution. The Shapiro normality test gives p-value 0.068 that indicates that the normality assumption is suitable for photosynthetic data at 0.01 and 0.05 level of significance. Also, the Bartlett test of homogeneity of variances gives p-value 0.30 that supports homogeneity of variances. Figure 2 shows the boxplot for square root transformation of photosynthetic rates data

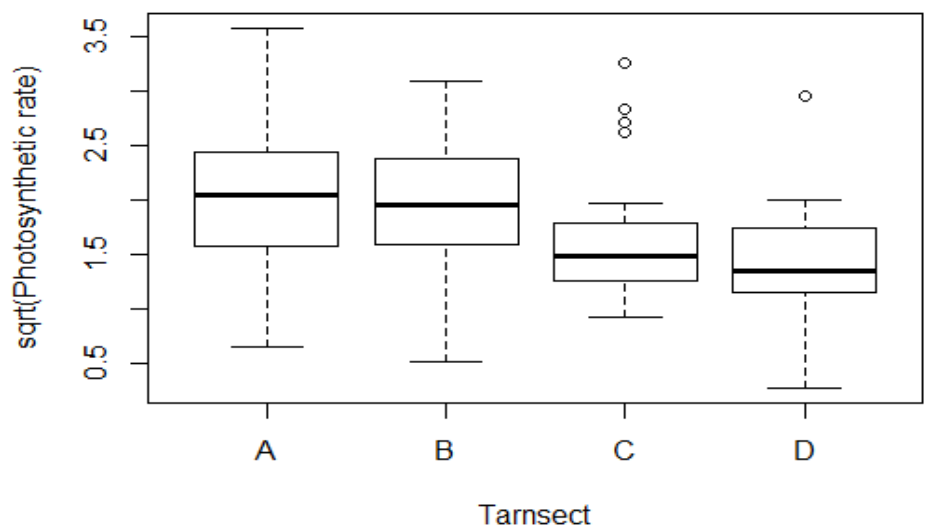

Figure 2 boxplots for square root transformation of photosynthetic rates data

### 4.1.1 ANOVA analysis

The analysis of variance for square root transformation of photosynthetic rates data is given in Table 2.

Table 2 ANOVA for square root of photosynthetic rates data

|            | Df  | Sum Sq | Mean Sq | F value | Pr(>F) |
|------------|-----|--------|---------|---------|--------|
| Treatments | 3   | 7.8    | 2.592   | 6.60    | 0.0004 |
| Residuals  | 104 | 40.8   | 0.393   |         |        |

Where the p-value in Table 2 is 0.0004, there are significance differences in group means at 0.05 but the ANOVA test does not show which group(s) is different.

### 4.1.2 gANOVA analysis ($1 - p_{adjusted}$ or $K_{adjusted}$ methods)

The gANOVA could be applied to square root transformation of photosynthetic rates data using alpha per family $1 - p_{adjusted}$ method or $K_{adjusted}$ method as described earlier. Note that the two methods must give the same conclusion.

**$1 - p_{adjusted}$ method**

In this method the groups are graphed against $1 - p_{adjusted}$ as

$g$ on $x$ axis  versus  $1 - p_{adjusted}$ on $y$ axis  with limits at $1 - \alpha(PF)$

The steps are

1. Use $n_T = 29 + 28 + 25 + 26 = 108$ and $G = 4$.
2. Compute $K_g$, $g = 1, 2, 3, 4$ using the data.
3. Find probabilities at $K_g$ from R-software (GB2 package) using
   $p = 1 - pgb2(K_g, 1, c(25.36, 25.68, 26.64, 26.32),\ 0.5,\ 52)$.
4. Obtain p-adjusted by using R-software function $padjust(p, method = "bh")$.
5. Graph $g$ against $1 - p_{adjusted}$
6. Graph the decision line at $DL = 1 - \alpha$
7. Any$(1 - p_{adjusted}) > 1 - \alpha$ reject $H_0$

Figure 3(a) shows the results of gANOVA using $\alpha = 0.05$. Where the $1 - P_{adjusted}$ values of groups A and D are outside the decision limit, the null hypothesis is rejected.

**$K_{adjusted}$ method**

In this method the groups are graphed against $K_{adjusted}$ as

$$g \text{ on } x \text{ axis versus } K_{adjusted} \text{ on } y \text{ axis with limits using } qGB2$$

The steps are

1. Use $n_T = 29 + 28 + 25 + 26 = 108$ and $G = 4$.
2. Compute $K_g$, $g = 1, 2, 3, 4$ using the data.
3. Find probabilities at $K_g$ from R-software (GB2 package) using

    $p = 1 - pgb2(K_g, 1, c(25.36, 25.68, 26.64, 26.32), 0.5, 52)$.
4. Obtain p-adjusted by using function $padjust(p, method = "bh")$.
5. Compute $K_{adjusted}$ from quantile function as

    $qgb2(1 - p_{adjusted}, 1, c(25.36, 25.68, 26.64, 26.32), 0.5, 52)$.
6. Find the decision limit using quantile function as

    $DL = qgb2(1 - \alpha, 1, c(25.36, 25.68, 26.64, 26.32), 0.5, 52)$.
7. Any$(K_{adjusted}) > DL$ reject $H_0$

Figure 3(b) shows the results of gANOVA using $\alpha = 0.05$. Where the $K_{adjusted}$ of groups A and D are outside the decision limit, the null hypothesis is rejected.

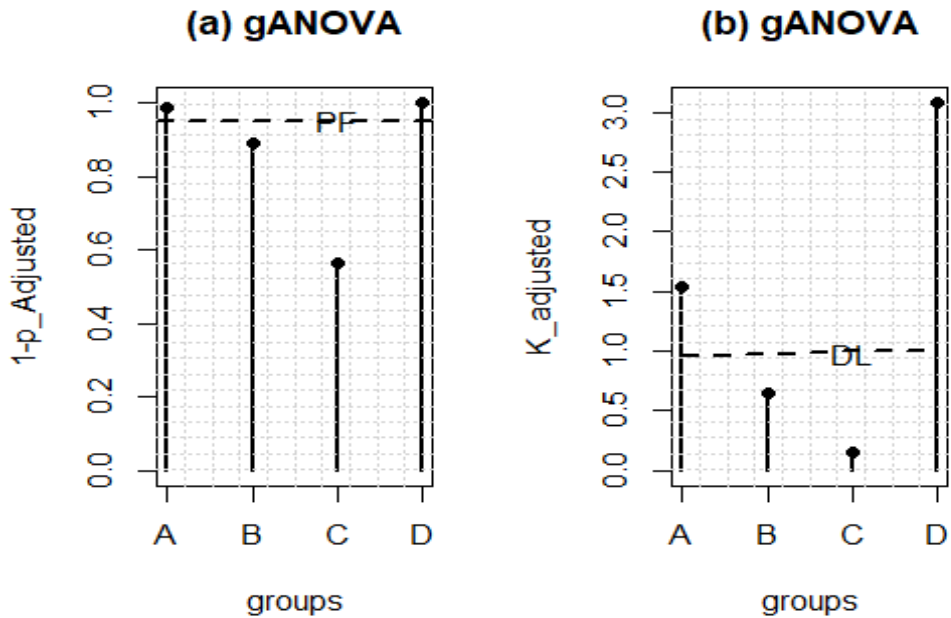

Figure 3 gANOVA for sqrt photosynthetic rates data using (a) $1 - p_{adjusted}$ method and (b) $K_{adjusted}$ method.

It is clear from Figure 3 gANOVA gives two conclusions:
1. $H_0$ is rejected
2. This rejection or difference is coming mainly from groups A and D.

On the other hand, one-way ANOVA in Table 2 is rejecting $H_0$ (0.0004<0.05) without showing which group(s) is causing the rejection. This required another investigation using Tukey honest significance differences test.

### 4.1.3  Tukey honest significance differences test

The Tukey honest significance differences test with 95% confidence interval is used with ANOVA and the results are given in Table 3.

Table 3 Tukey multiple comparisons of means with 95% family-wise confidence level

| Groups | Diff | Lwr | Upr | P_adj |
|---|---|---|---|---|
| B-A | -0.096 | -0.529 | 0.337 | 0.938 |
| C-A | -0.361 | -0.807 | 0.085 | 0.156 |
| D-A | -0.691 | -1.133 | -0.249 | 0.0005 |
| C-B | -0.265 | -0.715 | 0.185 | 0.419 |
| D-B | -0.595 | -1.041 | -0.149 | 0.0039 |
| D-C | -0.330 | -0.788 | 0.128 | 0.2426 |

Table 3 illustrates that
1. There are significance differences between averages for groups D-A and D-B.
2. There are no significance differences among remaining groups including A-B.

### 4.2  Application 2: simulated data

Four groups simulated data from normal distribution with means 105, 100, 98 and 103 and same variances is given in Table 4. The Shapiro normality test for this data gives p-value 0.5 that indicates that the normality assumption is suitable. Also, the Bartlett test of homogeneity of variances gives p-value 0.5 that supports homogeneity of variances.

Table 4. Simulation data from normal distribution with means 105, 100, 98 and 103 and equal variances

| A | B | C | D |
|---|---|---|---|
| 108.71 | 109.64 | 99.06 | 94.97 |
| 98.93 | 100.47 | 100.75 | 100.18 |
| 95.86 | 96.03 | 106.47 | 114.81 |
| 88.86 | 106.18 | 111.89 | 111.60 |
| 126.56 | 80.52 | 125.57 | 115.38 |
| 126.99 | 96.06 | 99.29 | 102.19 |
| 97.66 | 105.63 | 99.16 | 114.96 |
| 117.93 | 86.04 | 98.45 | 102.31 |
| 109.28 | 96.33 | 90.63 | 107.54 |
| 108.62 | 93.67 | 115.80 | 114.36 |
| 107.31 | 83.25 | 112.83 | 102.10 |
| 85.14 | 105.64 | 87.86 | 102.49 |
| 102.79 | 99.10 | 118.00 | 115.46 |
| 100.44 | 94.99 | 95.81 | 99.22 |
| 99.22 | 102.71 | 88.60 | 105.68 |

The ANOVA result for these simulated data is given in Table 5. Where the p-value is slightly more than 0.05, the null hypothesis of equal means could not be rejected at 0.05.

Table 5. ANOVA for simulated data

|  | Df | Sum Sq | Mean Sq | F value | Pr(>F) |
|---|---|---|---|---|---|
| Treatments | 3 | 809.82 | 269.94 | 2.71 | 0.0535 |
| Residuals | 56 | 5574.81 | 99.55 |  |  |

The Tukey honest significance differences test with 95% confidence interval is given in Table 6. With careful investigation, the results show different between groups A and B where the p-value for the comparison B-A is 0.048 < 0.05. In other words, the null hypothesis of equal means may be rejected at 0.05. This is different conclusion from ANOVA results in Table 5.

Table 6. Tukey multiple comparisons of means with 95% family-wise confidence level

| Groups | Diff | Lwr | Upr | P_adj |
|---|---|---|---|---|
| B-A | -7.9 | -17.51 | 1.8 | 0.048 |
| C-A | -1.6 | -11.25 | 8.0 | 0.97 |
| D-A | 1.9 | -7.72 | 11.6 | 0.95 |
| C-B | 6.3 | -3.39 | 15.9 | 0.32 |
| D-B | 9.8 | 0.15 | 19.4 | 0.15 |
| D-C | 3.5 | -6.11 | 13.2 | 0.77 |

On the other hand, Figure 4 shows the results of gANOVA for the simulated data using $K_{adjusted}$ method. Where the $K_{\text{adjusted}}$ for group B is outside the decision limit, the null hypothesis is rejected.

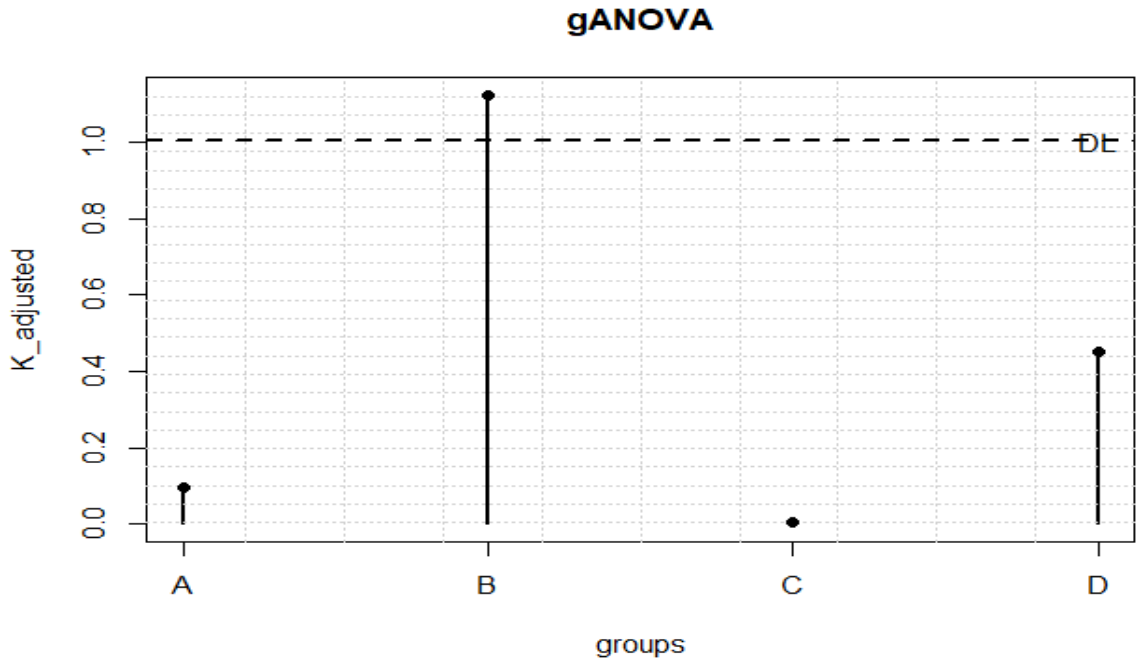

Figure 4 gANOVA for normal simulated data using $K_{adjusted}$ method

This application gives more insights into gANOV in comparison with ANOVA. While the ANOVA is not significance at 0.05, the Tukey honest significance differences test with 95% confidence interval has shown significance results at 0.05. On the other hand, gANOVA is showing significance results at 0.05.

## 5 Conclusion

A simultaneous test for means known as gANOVA is proposed as the ratio of between square for each treatment and within sum of squares for all treatments that created from F-test in one-way analysis of variance. The sum of this ratio is the F test in the analysis of variance.

The simulation results on the adjusted p-values are shown that the preferred method for gANOVA to control the type I error near to nominal value is Benjamini-Hochberg's method. The proposed method provides novel insights into the comparison among means where it collects the test and pairwise comparisons in one step.

The exact sampling distribution for the proposed method is derived as a beta distribution of the second type. Moreover, it may be considered gANOVA as an unblind test for F-test in one-way analysis of variance that gives more explanation and analysis for differences among treatment averages and identify which specific group mean is different simultaneously or graphically.

The proposed method is applied to photosynthetic rates of the oak seedlings data and simulated data from normal distribution. In simulation data application, an interesting result obtained. While the ANOVA is not significance at 0.05, the Tukey honest significance differences test with 95% confidence interval has shown significance results at 0.05 and this coincides with the significance results of gANOVA. Lastly, the extension of this method to other designs such as two-factor ANOVA needs more studies.

## References

[1] Abdi, H. (2007) The Bonferonni and Šidák corrections for multiple comparisons. In: Neil Salkind. Encyclopedia of Measurement and Statistics: Thousand Oaks (CA), Sage.

[2] Benjamini, Y., and Hochberg, T. (1995) Controlling the false discovery Rate: a practical and powerful approach to multiple testing. Journal of the Royal Statistical Society Series B, **85**, 289–300.

[3] Benjamini, Y. (2010) Simultaneous and selective inference: current successes and future challenges. Biometrical Journal, **52**, 708-721.

[4] Bretz, F., Maurer, W., and Hommel, G. (2011) Test and power considerations for multiple endpoint analyses using sequentially rejective graphical procedures. Statistics in Medicine, **30**, 1489-1501.

[5] Cochran, W. G., and Cox, G.M. (1957) Experimental Designs. 2nd edition. Wiley, New York.

[6] Coelho, C.A. and Mexia, J.T. (2007) On the distribution of the product and ratio of independent generalized gamma-ratio random variables. Sankhya: The Indian Journal of Statistics, **69**, 221-255.

[7] Dunn, O.J. (1964) Multiple comparisons using rank sums. Technometrics, **6**, 241-252.

[8] Fisher, R. A. (1918) The correlation between relatives on the supposition of mendelian Inheritance. Philosophical Transactions of the Royal Society of Edinburgh, **52**, 399–433.

[9] Fisher, R. A., (1925) Statistical methods for research workers. Edinburgh: Oliver and Boyd.

[10] Diaz-Garcia, J., and Jaimez, R. (2010) Bimatrix variate generalized beta distributions: theory and methods. South Africa Statistical Journal, **44**, 193-208.

[11] Elamir, E. H. (2012) On uses of mean absolute deviation: decomposition, skewness and correlation coefficients. Metron: International Journal of Statistics, **70**, 145-164.

[12] Hochberg, Y. (1988) A sharper Bonferroni procedure for multiple tests of significance. Biometrika, **75**, 800–802.

[13] Holm, S., (1979) A simple sequential rejective multiple test procedure. Scandinavian Journal of Statistics, **6**, 65–70.

[14] Hommel, G. (1988) A stagewise rejective multiple test procedure on a modified Bonferroni test. Biometrika, **75**, 383 – 386.

[15] Hommel, G. (1989) A comparison of two modified Bonferonii procedures. Biometrika, **76**, 624-625.

[16] Kutner, M., Nachtsheim, C., Neter, J. and William, L. (2004) Applied linear statistical models. 5th Ed., McGraw-Hill/Irwin.

[17] Montgomery, D.C. (2013) Design and analysis of experiments. 8th ed. Jhon Wiley & Sons. Inc.

[18] Sidak, Z. (1967) Rectangular confidence regions for the means of multivariate normal distributions. Journal of the American Statistical Association, **62**, 626-633.

[19] Simes, R. J. (1986) An improved Bonferroni procedure for multiple tests of significance. Biometrika, **73**, 751-754.

[20] Westfall, P. (2005) Combining p-values. in Encyclopedia of Biostatistics, eds. P. Armitage and T. Colton, Chichester: Wiley, pp. 987-991.

**Appendix:** R program for gANOVA

```r
library(matrixStats)
library(moments)
library(GB2)
gANOVA = function(x,ng,g,aa) { # x:data matrix, ng: group size, g # group, aa: alpha
 y0 = matrix(x,ng,g)   ## data
 y1 = ifelse(y0=="NA",0,1) ## # of values
 y2 = colSums(y1,na.rm=T) # # of ng each group
 nt = sum(y2)          ## nt total size
 m0 = mean(y0,na.rm=T)       ## overall mean
 vx = var(x,na.rm=T)    ## var. all data
 r0 = matrix(m0,ng,g)
 ## Exact between
 BDm = colMeans(y0,na.rm=T)     ## group means
 BDm0 = matrix(BDm,ng,g,byrow=T)   ## rep BDm
 BD0 = (BDm0-r0)^2  ## between square
 BD1 = y1*BD0     ### to get na location
 BD2 = colSums(BD1,na.rm=T) ## col sum between
 BD3 = sum(BD2)  ## Exact Between
 ##Exact within ###
 WD0 = (y0-BDm0)^2  ## within square
 WD1 = sum(WD0,na.rm=T)   ## exact within
 H0 = (BD2/(g-1))/(WD1/(nt-g))
 q0 = qgb2(1-aa/g,1,(nt-g)/g,0.5,(nt-g)/2)
 p1 = pgb2(H0,1,(nt-g)/g,0.5,(nt-g)/2)
 p0 = 1-p1;
 bon = p.adjust(p0, method="bonferroni")
bh = p.adjust(p0, method="BH")
 bh0 = 1-bh
 q00 = qgb2(1-aa,1,(nt-g)/g,0.5,(nt-g)/2)
 F0 = function(y,g,nt) {qgb2(y,1,(nt-g)/g,0.5,(nt-g)/2)}
 F00 = sapply(bh0,F0,g=g,nt=nt)
## graph
 par(mfrow=c(1,2))
## first method
```

```r
  plot(1:g,1-bh,type="h",xaxt="n",lwd=2,ylim=c(0,1),
  xlab = "groups",ylab="1-p_Adjusted",main="(a) gANOVA")
  axis(1,at=1:g,labels=LETTERS[1:g])
  grid(10,25)
  points(1:g,1-bh,pch=16)
  text(g-1,1-aa,"PF")
  abline(h=1-aa,lwd=2,lty=2)
#Second method
plot(1:g,F00,type="h",xaxt="n",lwd=2,ylim=c(0,max(q00,F00)),
  xlab="groups",ylab="K_adjusted",main="(b) gANOVA")
  axis(1,at=1:g,labels=LETTERS[1:g])
  grid(10,25 )
  points(1:g,F00,pch=16)
  text(g-1,q00,"DL")
  abline(h=q00,lwd=2,lty=2)
}
```